\documentstyle[epsfig]{aipproc}
\begin{document}

\author{V.I. Dokuchaev$^*$\thanks{E-mail ID: dokuchaev@inr.npd.ac.ru},
Yu.N. Eroshenko$^*$, L.M. Ozernoy$^{\dagger\ddagger}$\thanks{E-mail ID:
ozernoy@science.gmu.edu, ozernoy@stars.gsfc.nasa.gov}}

\address{$^*$Institute for Nuclear Research of the
Russian Academy of Sciences, \\ 117312 Moscow, Russia\\
$^{\dagger}$George Mason University, Fairfax, VA 22030-4444, USA\\
$^{\ddagger}$Code 685, Laboratory for Astronomy and Solar Physics,
\\ NASA/Goddard Space Flight Center, Greenbelt, MD 20771, USA}

\title{The Eddington Luminosity Phase \\
in Quasars: Duration and Implications}

\maketitle

\def\lax    {\ifmmode{_<\atop^{\sim}}\else{${_<\atop^{\sim}}$}\fi}
\def\gax    {\ifmmode{_>\atop^{\sim}}\else{${_>\atop^{\sim}}$}\fi}
\begin{abstract}
Non-steady and eruptive phenomena in quasars are thought to be associated
with the Eddington or super-Eddington luminous stage. Although there is no
lack in hypotheses about the total duration of such a stage, the latter
remains essentially unknown.
We calculate the duration of quasar luminous phase in dependence upon the
initial mass of a newborn massive black hole (MBH) by comparing the
observed luminosity- and redshift distributions of quasars with
mass distribution of the central MBHs in normal galactic nuclei. It is 
assumed that, at the quasar stage, each MBH goes through a single (or
recurrent) phase(s) of accretion with, or close to, the Eddington
luminosity. The mass distributions of quasars is found to be
connected with that of MBHs residing in normal galaxies by a
one-to-one corrrespondence through the entire mass range of the
inferred MBHs if the accretion efficiency of mass-to-energy
transformation $\eta\sim 0.1$.

\end{abstract}

\section*{Introduction}
An approximate relationship $M_h\simeq (0.003-.006)M_b$ between
the central MBH mass, $M_h$, and that of the galactic bulge, $M_b$,
has been established for a few dozen of galaxies, both nearby and more 
distant ones \cite{0003},\cite{0006}. A relationship between absolute
magnitudes of quasars and their host galaxies found in \cite{Bahc}
is reduced to the MBH to bulge mass relation in galaxies provided
that \cite{oz98}: (i) the central MBH shines at or near to the
Eddington luminosity and (ii) the host galaxy undergoes through a
starburst episode. This correlation, coupled with the known
luminosity function of galaxies, can serve to obtain \cite{S98} the
MBH mass distribution $\phi_1(M_h)dM_h$. The history of matter
accretion onto a central MBH thought to serve as a source of quasar
activity is linked to the present observable properties of each
individual quasar, such as its luminosity, variability, and
emission spectrum. If the bolometric luminosity of a quasar
comprises a fraction $\lambda$ of the Eddington luminosity, i.e.
$\lambda=L/L_E$, $L_E=4\pi GM_hm_pc/ \sigma_T$, the underlying
accretion is accompanied by an exponential growth of the MBH mass
with the characteristic time
$t_{E}=4.5\cdot10^8\eta/\lambda\mbox{~yrs}$, where $\eta$ is the
accretion efficiency of mass-to-energy transformation. The crucial
problem is the duration, $t_q$, of such a nearly Eddington
accretion phase. Usually an effective $t_q$, the same for the
entire black hole mass range $M_h\sim10^6 - 10^{10}M_\odot$, is
calculated by comparing the global number density of normal
galaxies and quasars and is found to be
$t_q=10^6-5\cdot10^8$~years in \cite{HNR97}.
Meanwhile the recent data on mass distribution of MBHs in galaxies
provide an opportunity to solve this problem in a more detailed
way, {\it viz.,} to calculate the dependence of $t_q$ upon $M_h$,
which is the major aim of this paper. We shall explore whether the
distribution functions of quasars and MBHs in normal galaxies are
consistent with each other, and we will do this locally in the
vicinity of each mass.

\section*{The Eddington Luminosity Phase}
\label{eddington}

It would be reasonable to assume that the duration of the Eddington
phase $t_q$ depends on the initial mass of a newborn MBH or, in
other words, on the initial luminosity, $L_i$, of the quasar:
$t_q=t_q(L_i)$. For simplicity, the transition to
and out of the Eddington phase is supposed to occur instantaneously:
\begin{equation}
L=\left\{ \begin{array}{ll} 0, &  \mbox{ if  } t<t_i; \label{lll} \\
L_i\exp[(t-t_i)/t_{E}], &  \mbox{ if  } t_i<t<t_i+t_q(L_i); \\
0, &  \mbox{ if  } t>t_i+t_q(L_i),
\end{array}
\right.
\end{equation}
where $t_i$ is the instant of the MBH formation. Along with the distribution 
function of MBHs in the galactic nuclei, $\phi_1$,
we use the observed distribution of quasars in absolute magnitude $M_B$ and
redshift $z\le z_e\sim3$, $\phi_2(M_B,z)dM_Bdz$, taken from
\cite{boyle91}. The balance equation is given by
\begin{equation}
\frac{2{.}5}{\ln10}
\int\limits_0^{\infty}dz(1+z)^{-3/2} \phi_2(M_B(L),z)
\simeq\frac{t_E(L)}{t_{0}} \int\limits_{M(L)}^X\phi_1(M_h)dM_h,
\label{intint}
\end{equation}
where $X=M(L)\exp[t_q(L)/t_{E}]$ with $M(L)$ determined by equation
$L=\lambda L_E$ and a relationship $t=t_0/(1+z)^{3/2}$ for the flat
cosmological model is used. While obtaining Eq.~(\ref{intint}), we
have also taken into account that $t_E$ very slowly varies with
respect to $\phi_1(M_h)$.  Eq.~(\ref{intint}) defines $t_q$ as an
implicit function of $L_i$, which can be translated into a
relationship between $t_q$ and the initial BH mass $M_i$ by using
equation $L=\lambda L_E$.

\section*{Results}
Numerical solution of Eq.~(\ref{intint}) is found by adopting the
mass distribution of MBHs $\phi_1(M_h)dM_h$ from \cite{S98},
derived with the use of three relationships, {\it viz.,} (i) a
correlation $\log(M_h)=\log(M_b)-2.6\pm0.3$ between
the MBH mass $M_h$ and the bulge mass $M_b$; (ii) the
mass-luminosity relation for galaxies, and (iii) the Schechter
luminosity function.
\begin{figure}[t!] 
\centerline{\epsfig{file=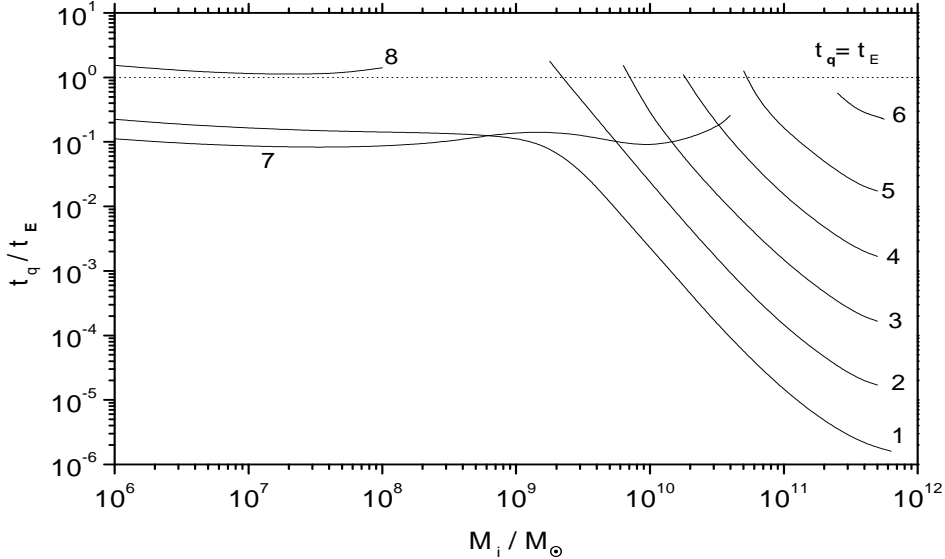,height=3.35in,width=5.5in}}
\caption[fig1]{The ratio  $t_q(M_i)/t_E$ as a function of the
initial MBH mass $M_i$. Curves labeled 1 to 6 are based on 
distribution A for $\eta=$$10^{-1}$, $10^{-2}$, $10^{-3}$,
$10^{-4}$, $10^{-5}$, and $10^{-6}$, respectively. Curves 7 and 8
are based on distribution B for $\eta=$$10^{-1}$ and $10^{-2}$,
respectively.
\label{Fig.1} }
\end{figure}
In Fig.~1 and Fig.~2, we employ two somewhat different distributions
in MBH mass from \cite{S98} and name them `distribution A' 
and `distribution B', 
which correspond to the power-law and log-Gaussian shape of dispersion,
respectively.
Fig.~1 presents the results of our numerical computation of the
ratio $t_q(M_i)/t_E$ for different values of $\eta$ and MBH mass
distributions A and B. It should be noted that if $\eta<0.1$, the
solution exists not for all values of $M_i$. The domain where the solution
exists is determined from the condition that the r.h.s of
Eq.~(\ref{intint}) exceeds its l.h.s. if one puts $X=+\infty$. For
those $M_i$ which lead to an opposite condition, the number of
galactic nuclei with MBHs is not enough to explain, in the
framework of our model, the distribution function of quasars in
$M_B$ and $z$, even if these MBHs stay in an active quasar state
during the maximum possible time $t_q\sim3t_E$. The solution only
exists at $\eta>7\cdot10^{-7}$ for BH mass distribution A and at
$\eta>6\cdot10^{-3}$ for distribution B. A single-valued mapping
$M_i\to M_h$ breaks up on the left end of curves 2 to 6. 

Fig.~1 demonstrates the main result of this work:
{\it distributions of MBHs and quasars in mass are connected
by one-to-one correspondence through the whole range of the observed
masses only for $\eta\sim 0.1$}, both for the distribution A (curve
1) and B (curve 7). This concordance breaks down for a certain
range of MBH masses, {\em viz.\/} the solution of Eq.~(\ref{intint})
does not exist if $\eta\ll 0.1$. Nevertheless, the jumps of the
$\eta$ value are not excluded on the boundary of the domain, where the
solution exists. Similar jumps seem to be quite natural if MBHs
in the different mass ranges are formed by different ways (e.g., by
collapse of massive gas clouds, stellar clusters, etc., see review
by Rees 1984) and so there are various accretion regimes with
different values of $\eta$. If such jumps indeed take place,
transitions between the curves of each of distributions A and B are
possible. These transitions must be smoothed because the MBHs formed by
different ways would coexist in some mass range(s). The {\it most
probable value of $\eta$ established above is} $\eta\sim0.1$, which
corresponds to curves 1 and 7 in Fig.~1. For both these curves, the
relationship $t_q<t_E$ is carried out and therefore the BH mass
growth is not substantial -- it generally does not exceed a value
comparable to the initial BH mass, and for $M_h>10^7~{\rm M}_\odot$
it is negligible.
\begin{figure}[t!] 
\centerline{\epsfig{file=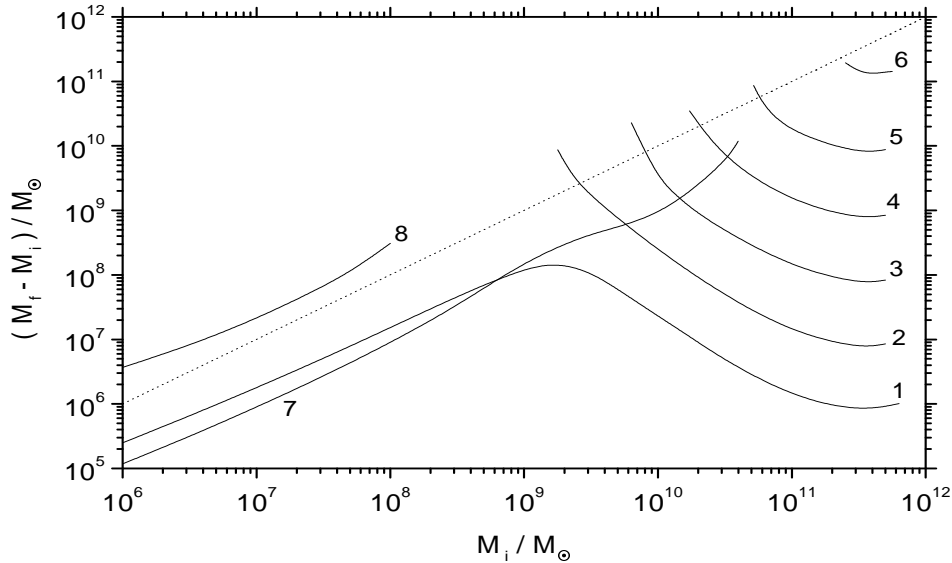,height=3.35in,width=5.5in}}
\caption{The accreted mass,  $M_f-M_i$, as a function of the initial mass
$M_i$. The curves are labeled in the same way as in Fig.~1.
\label{Fig.2} }
\end{figure}

\end{document}